\documentclass[pdflatex,sn-nature,twocolumn]{sn-jnl}

\usepackage{graphicx}%
\usepackage{multirow}%
\usepackage{amsmath,amssymb,amsfonts}%
\usepackage{amsthm}%
\usepackage{mathrsfs}%
\usepackage[title]{appendix}%
\usepackage{xcolor}%
\usepackage{textcomp}%
\usepackage{manyfoot}%
\usepackage{booktabs}%
\usepackage{algorithm}%
\usepackage{algorithmicx}%
\usepackage{algpseudocode}%
\usepackage{listings}%
\usepackage[switch]{lineno}

\unnumbered

\begin{document}


\makeatletter
\newcommand{\manuallabel}[2]{\def\@currentlabel{#2}\label{#1}}
\makeatother

\manuallabel{SFig:pillars}{S1}
\manuallabel{SFig:cagedATP}{S2}
\manuallabel{SFig:velocity_and_thickness}{S3}
\manuallabel{SFig:spacing}{S4}
\manuallabel{SFig:fully_developed}{S5}
\manuallabel{SFig:pillars_feathers}{S6}
\manuallabel{SFig:nocurvature}{S7}
\manuallabel{SFig:nofeathers}{S8}
\manuallabel{SFig:vcorr}{S9}

\manuallabel{SMov:experiment}{S1}
\manuallabel{SMov:pillars}{S2}
\manuallabel{SMov:cagedATP}{S3}
\manuallabel{SMov:transversal_oscillations}{S4}
\manuallabel{SMov:transition_feathers}{S5}
\manuallabel{SMov:feather_formation}{S6}
\manuallabel{SMov:nocurvature}{S7}
\manuallabel{SMov:ModelA}{S8}
\manuallabel{SMov:ModelB}{S9}


\title[Article Title]{Spatiotemporal crystallization of an active fluid}


\author[1,2]{\fnm{Olga} \sur{Bantysh}}\email{olga.b.bantysh@ub.edu}

\author[1,3]{\fnm{Ramon} \sur{Reigada}}\email{reigada@ub.edu}

\author[4]{\fnm{Rodrigo C. V.} \sur{Coelho}}\email{rcvcoelho@cbpf.br }

\author[5]{\fnm{Pau} \sur{Guillamat}}\email{pguillamat@ibecbarcelona.eu}

\author*[1,2]{\fnm{Jordi} \sur{Ignés-Mullol}}\email{jignes@ub.edu }
\author[1,2]{\fnm{Francesc} \sur{Sagués}}\email{f.sagues@ub.edu }

\affil*[1]{\orgdiv{Department of Materials Science and Physical Chemistry}, \orgname{Universitat de Barcelona}, \orgaddress{\street{Martí i Franquès 1}, \city{Barcelona}, \postcode{08028}, \state{Catalonia}, \country{Spain}}}

\affil[2]{\orgdiv{Institute of Nanoscience and Nanotechnology}, \orgname{University of Barcelona}, \orgaddress{\street{Martí i Franquès 1}, \city{Barcelona}, \postcode{08028}, \state{Catalonia}, \country{Spain}}}

\affil[3]{\orgdiv{Institute of Theoretical and Computational Chemistry (IQTCUB)}, \orgname{University of Barcelona}, \orgaddress{\street{Martí i Franquès 1}, \city{Barcelona}, \postcode{08028}, \state{Catalonia}, \country{Spain}}}

\affil[4]{\orgname{Centro Brasileiro de Pesquisas Físicas}, \orgaddress{\street{Rua Xavier Sigaud 150}, \city{Rio de Janeiro}, \postcode{22290-180}, \state{RJ}, \country{Brazil}}}

\affil[5]{\orgname{Institute for Bioengineering of Catalonia (IBEC)}, \orgaddress{\street{Baldiri Reixac, 10}, \city{Barcelona}, \postcode{08028}, \state{Catalonia}, \country{Spain}}}


\abstract{The emergence of long-range spatiotemporal order from intrinsic chaos is a central challenge in far-from-equilibrium physics. In active fluids, such as cytoskeletal networks driving cellular motion, self-generated flows typically produce “active turbulence,” lacking translational symmetry. Here we show that a chaotic active nematic can self-organize into a spatiotemporal crystal, forming a regular lattice of density, orientation, and vorticity that breaks both spatial and temporal translational symmetry. Using a microtubule/kinesin active nematic interfaced with a lamellar liquid crystal and confined in microfluidic channels, we observe robust spatiotemporal lattices without external forcing. The ordering emerges from spontaneous synchronization of intrinsic flow instabilities, mediated by confinement and feedback between the active layer and the passive anisotropic interface. Continuum nematohydrodynamics simulations support our interpretation, highlighting how intrinsic length and time scales shape the active crystals. These results reconcile chaos and crystallinity in active matter and provide a strategy for engineering order in self-driven, far-from-equilibrium soft materials.}




\maketitle

\section{Introduction}\label{sec:intro}
Two of the most disparate concepts in physics are those of order and chaos. Following Strogatz, we could bridge them by referring to the phenomenon of synchrony. In fact, in relation to his studies on nonlinear dynamics and chaos, he suggested this idea with elegant words “...synchrony is perhaps the most pervasive and mysterious drive in Nature” \cite{Strogatz2003}.  We demonstrate here this vision by reporting a pattern-forming mechanism in a chaotic regime of an active fluid. The latter refer to fluids characterized by their ability to self-organize intricate flows, deprived of any spatio-temporal regularity \cite{Wensink2012,giomi2015,Alert2022,Urzay2017,Martinez-Prat2021,Tan2019}.

Paradigmatic patterns in nonlinear physics, such as Faraday instabilities, Turing patterns, and Rayleigh-Benard flows, exhibit intrinsic length scales \cite{Cross1993}. Patterned hydrodynamic currents arise from an externally driven source that destabilizes reference states that are at rest and spatially uniform. In this work, we challenge this principle with an scenario of fluid pattern formation that emerges from a turbulent, time evolving regime in an active nematic (AN) fluid, characterized by its long-range orientational order \cite{doostmohammadi2018active}. The formation of a spatiotemporal lattice occurs with no external actuation, instead leveraging the internal stresses generated at the component scale.

From an experimental point of view, a paradigmatic realization of AN is a layer of fluorescently-labeled microtubules powered by adenosine triphosphate (ATP)-fed kinesin clusters, condensed at the aqueous-oil interface. The self-propelled filaments organize in dense bundles that form textures that are continuously reconfigured while stretched and ruptured under a kind of perpetual and unpredictable motion \cite{Sanchez2012}.

\begin{figure*}[t]
\centering
\includegraphics[width=\textwidth]{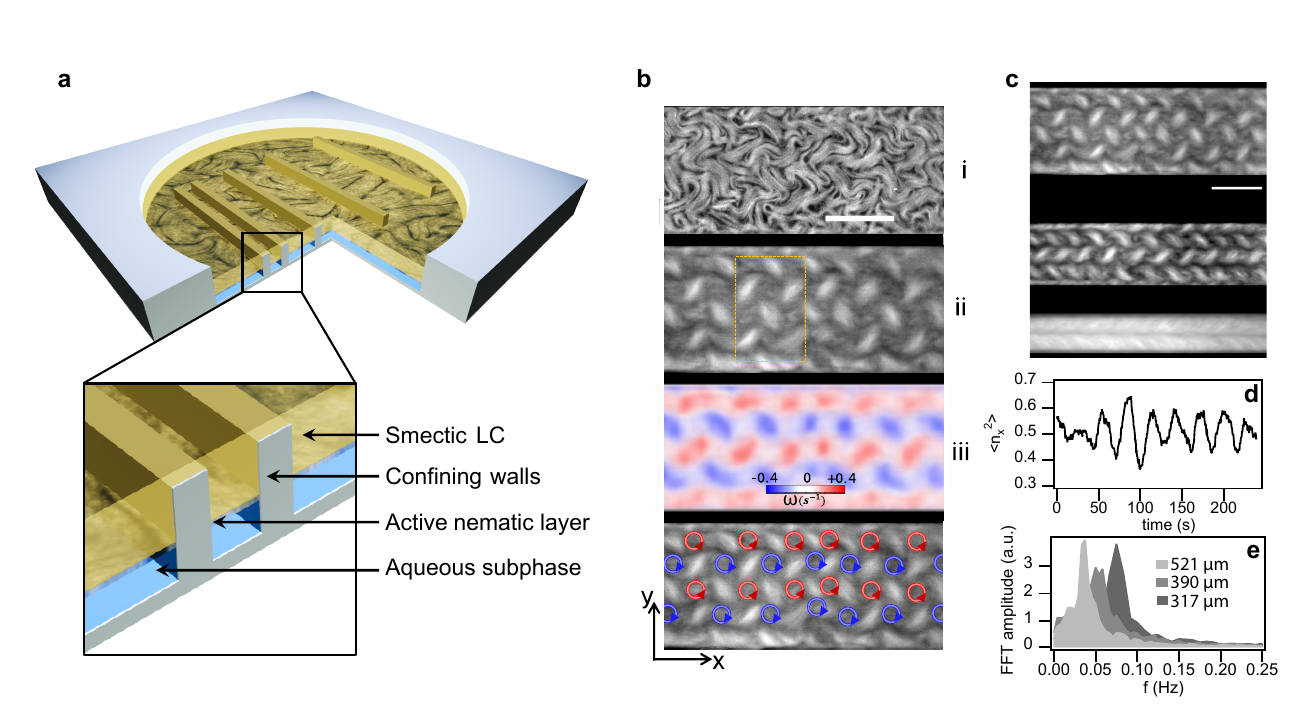}
\caption{Experimental setup and self-organization of active flows.  \textbf{a} Top view  of experimental confined sample in an open cell. Lateral confinement is imposed with a series of polymeric walls. A layer of liquid crystal oil (8CB) is placed on top of the aqueous subphase. The AN layer forms at the water/oil interface. \textbf{b} Self-organization of the AN in density hot-spots and vortex lattices. ATP concentration is 1.37 mM. Scale bar, 200 $\mu$m. i) Fluorescence microscopy snapshot. ii) Time average of 240 fluorescence images obtained at a rate of 2 fps. iii) Map with the average vorticity of the active flow field. iv) Image in ii with a sketch of the vortex lattice shown in iii overlaid. \textbf{c} Time average of 240 fluorescence micrographs of the AN acquired at 2 fps. Three channels of different widths are observed simultaneously. Scale bar, 200 $\mu$m. \textbf{d} Time dependence of the space-averaged longitudinal component of the director field, $<n_x^2>$, for a channel of width 521 $\mu$m. The size of the averaging window is shown as an orange frame in panel \textbf{b}.ii. \textbf{e} Average Fourier transforms of time traces of $<n_x^2>$ for channels of different width, as specified.}
\label{Fig:setup}
\end{figure*}

Some features of these disordered flows have been partially controlled in recent years, such as their preferred direction of self-propulsion or the distribution of active vortex sizes, using either lateral confinement \cite{Shendruk2017,Hardouin2019,Tan2019,opathalage19} or interfacing the active layer with anisotropic oils \cite{guillamat2016,Bantysh2024}. Nevertheless, the iconic chaotic dynamics in AN seems to preclude the formation of spatiotemporal patterns that are regular both in space and in time. While vortex lattices in AN have been predicted in simulations at the crossover between hydrodynamics-dominated and friction-dominated regimes \cite{Doostmohammadi2016}, no spatiotemporal crystallization has ever been reported.

In this work, we present experiments, supported by numerical simulations, showing that a combination of anisotropy and confinement leads to the emergence of spatiotemporal crystals \cite{Kongkhambut2022,Zhao2025} in ANs. The viscous anisotropy at the water/oil interface transforms the naturally chaotic AN currents into quasi-laminar flows \cite{guillamat2016}. At the same time, microfluidic confinement and a feedback mechanism with the passive fluid layer synchronizes the flow instabilities inherent to the extensile active filaments \cite{Thampi2014}. As a result, the AN system is organized as a spatiotemporal antiferromagnetic lattice of flow vortices concurrent with a herringbone pattern of the orientational and density fields. Interestingly, this synchrony reveals the role of the intrinsic length and time scales that set the transversal and longitudinal periodicity of the spatiotemporal active crystal. 

\section{Results}\label{sec:results}

\subsection{Emergence of a spatiotemporal crystal}

In our experiments, we interface the quasi-two-dimensional microtubules/kinesin AN layer with the liquid crystal oil octyl cyanobiphenyl (8CB), which features a highly anisotropic lamellar phase (Smectic-A, SmA) in the biocompatible temperature range $21.5\,^{\circ}$C - $33.5\,^{\circ}$C (see Methods section). 8CB molecules align  in the presence of a magnetic field, creating two directions with giant viscosity contrast. As a result, the naturally chaotic currents of the AN, which are hydrodynamically coupled to the oil, transform into an array of anti-parallel quasi-laminar flow lanes perpendicular to the applied magnetic field 
\cite{guillamat2016}. The flow lanes concentrate the self-propelled positive half-integer defects of the nematic textures (comet-shaped folds of the filament bundles). 
The spacing between neighboring flow lanes, $\ell_a$, is proportional to the intrinsic length scale of the AN, $\sqrt{K/\zeta}$~\cite{PhysRevLett.89.058101,guillamat2016,Calderer2025}, at which elastic and active stresses balance \cite{giomi2015}. Here, $K$ is an effective elastic constant of the AN, and $\zeta$ is the activity parameter \cite{doostmohammadi2018active}.

The dynamics of the active fluid is changed when the same preparation is conducted imposing both viscous anisotropy and lateral confinement (see Fig. \ref{Fig:setup}a and Methods section). Under high activity conditions and for channels similar to but wider than  $\ell_a$, which is measured outside the channels, the quasi-laminar aligned flow is disturbed. The system seemingly reverts to the isotropic turbulent regime, as seen both in the AN fluorescence textures and flow (see Fig. \ref{Fig:setup}b.i and Movie \ref{SMov:experiment}). The fluorescence intensity in the images is proportional to the local density of active filaments, which are interspaced by the voids where defect cores reside. A time average of a sequence of fluorescence micrographs, which would result in a homogeneous density map for isotropic turbulence, reveals here the emergence of a herringbone lattice of regions with a higher probability for the presence of MT bundles (henceforth referred to as "hotspots"), see Fig. \ref{Fig:setup}b.ii. Conversely, the dark areas between density hotspots have a higher probability for the passage of defects. Indeed, a time average of the vorticity computed from the measured velocity field (see Methods) reveals an antiferromagnetic vortex lattice (Fig. \ref{Fig:setup}b.iii) that is complementary to the density lattice (Fig. \ref{Fig:setup}b.iv), as active flows are more intense in the space between density hotspots.

These phenomena occur in a limited range of channel widths. When the channel is much wider than $\ell_a$, the system recovers the quasi-laminar regime of aligned flows. We can visually assess this result by performing the experiment in a system where a broad range of lateral confinements coexist, such as when the fluids are embedded in an array or cylindrical pillars (Fig. \ref{SFig:pillars} and Movie \ref{SMov:pillars}). In this case, density and vorticity hotspots only emerge between nearest neighbor pillars. In the usual experiments with the channel geometry, spatiotemporal crystallization does not occur inside channels narrower than $\sim 3\ell_a$. In Fig. \ref{Fig:setup}c we show the average fluorescence map of AN confined within channels of different widths. We see that the hotspot pattern does not form inside the narrowest channel, of width slightly lower than $3\ell_a$ ($\ell_a = 74\pm4 \mu$m in these experiments).

Besides the crystalline arrangement that becomes evident after time averaging, we have observed that these patterns feature intrinsic temporal periodicity. In Fig. \ref{Fig:setup}d, we report the temporal evolution of the squared longitudinal component of the AN orientational field, $<n_x^2>$, in a channel of width 521 $\mu$m, averaged over a window that includes the region that is in perfect focus in the transversal direction and encompasses two hotspots in the longitudinal direction (see rectangular frame in Fig. \ref{Fig:setup}b.ii). We have verified that oscillations are coherent within windows this size. This measurement is performed by sliding the window along the channel, and averaging the amplitude of the Fourier transforms computed in each window. The result is shown in Fig. \ref{Fig:setup}e for channels of different width, evidencing an intrinsic periodicity that changes with channel width. 

Finally, we have observed that the spatiotemporal crystal only emerges for high enough activities. We have put this into evidence by gradually increasing the ATP concentration in the system using caged ATP, a molecule that releases the chemical fuel upon UV light irradiation, as seen in Movie \ref{SMov:cagedATP} (see Methods). At low activities, flows are chaotic. At intermediate activities, we obtain quasi-laminar flows, but without hotspots. At high activities, quasi-laminar flows disappear and spatiotemporal active crystals emerge (Fig. \ref{SFig:cagedATP}).  

\begin{figure}[t]
\centering
\includegraphics[width=\columnwidth]{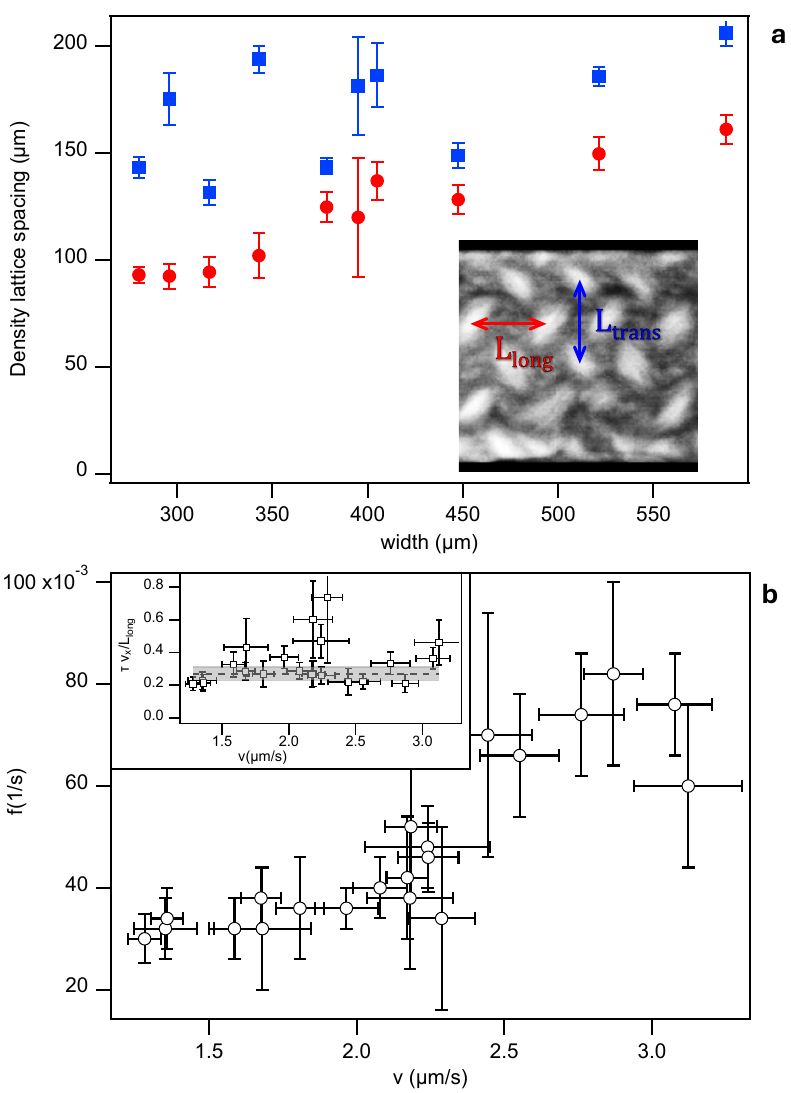}
\caption{Intrinsic scales of the spatiotemporal patterns. \textbf{a} Transversal, $L_{trans} (\square)$, and longitudinal, $L_{long} (\circ)$, lattice parameters of the density hot-spot lattice as a function of channel width. Error bars are the standard deviation from the mean in repetitions. \textbf{b} Characteristic oscillation frequency of the orientational field in the spatiotemporal crystals as a function of the average speed, used as a proxy for activity. In the inset, the scaling proposal $L_{long} \sim \tau\,v_x$ is compared with the experimental value of $L_{long}$. The shaded region is the 99\% confidence band of the weighted mean. Error bars indicate the 99\% confidence interval of the parameters.}
\label{Fig:length_scales}
\end{figure}
\begin{figure*}[thb]
\includegraphics[width=\textwidth]{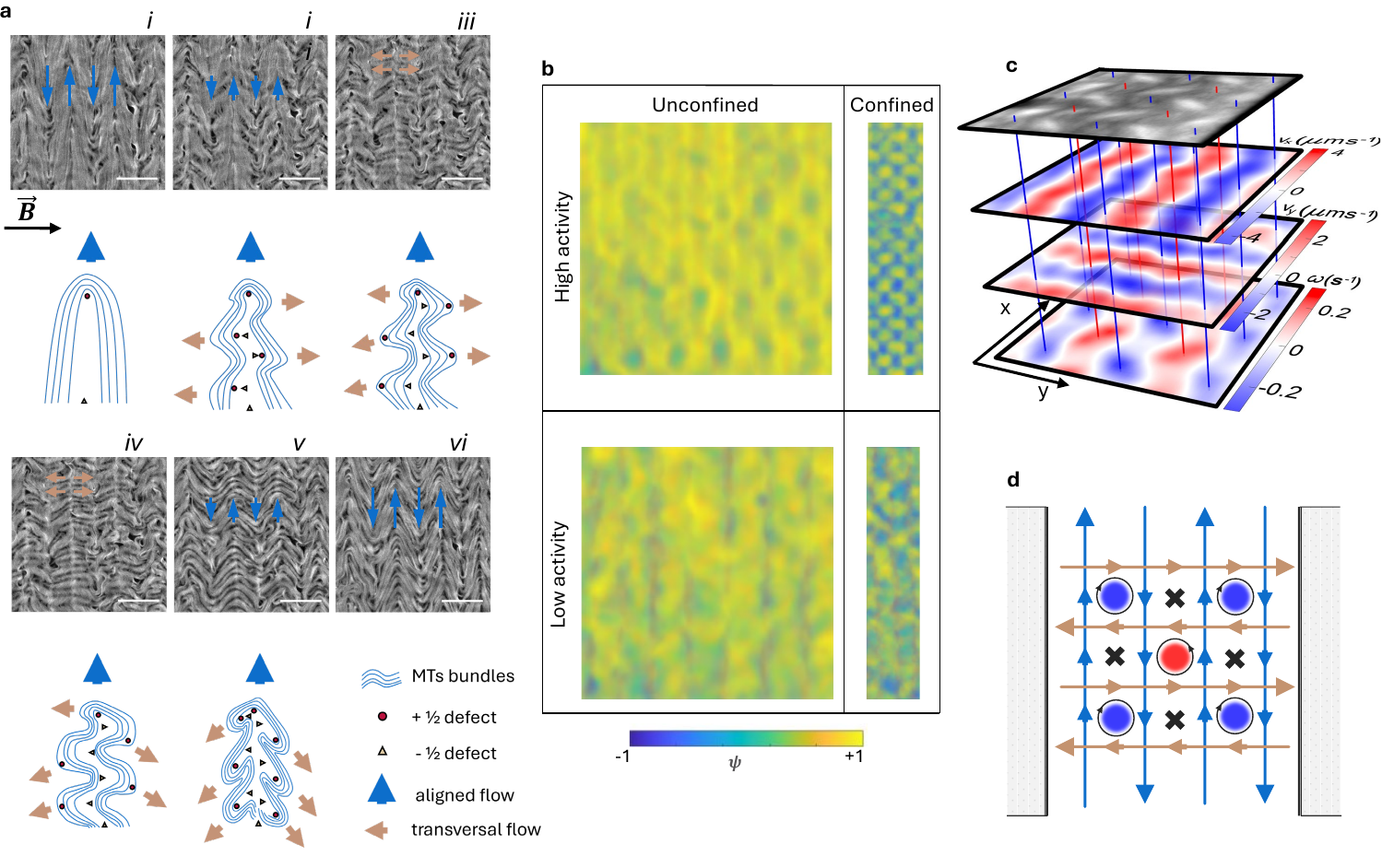}
\caption{Synchronized transversal instabilities trigger the formation of spatiotemporal crystals. \textbf{a}  Fluorescence micrographs showing the self-organization of transversal instabilities of the antiparallel aligned flows in an unconfined AN layer. Elapsed times from the leftmost frame, and then top to bottom, are 4 s, 6 s, 8 s, 16 s, and 22 s (see also Movie \ref{SMov:transversal_oscillations}). Scale bars, 200 $\mu$m. The orientation of the magnetic field is indicated below the first frame. A sketch of the evolution of active filaments and topological defects is depicted below the images. \textbf{b} Map of the time averaged order parameter of the AN orientational field ($\psi = <2\sin^2(\theta)-1>$, under different conditions. Here, $\theta$ is the angle between the local director field of the AN and $\vec{B}$, and $<...>$ means time average). \textbf{c} Overlay, from bottom to top, of the time-averaged (479 frames at 2 fps) vorticity, transversal velocity, longitudinal velocity, and fluorescence image of a region within a channel of width 430 $\mu$m. Red and blue vertical rods are guides for the eye to highlight the position of vortices in the other images. \textbf{d} Sketch illustrating the synchronization of antiparallel aligned flows (blue arrows) and transversal flows (brown arrows) for the confined AN. Complementary lattices of vortices (blue and red circles) and stagnation regions (black crosses) emerge. 
}
\label{Fig:transversal}
\end{figure*}

\subsection{Spatiotemporal crystal lattice parameters}

We next characterize the emergent lattices of MT density hotspots and active flow vortices. To this purpose, we define the spacing along the channel, \emph{i.e.}, the longitudinal lattice parameter, $L_{\text{long}}$, and the spacing perpendicular to the channel, \emph{i.e.}, the transversal lattice parameter, $L_{\text{trans}}$. We have measured these length scales in experiments spanning the range of channel widths at which the emergent lattices are stable (Fig. \ref{Fig:length_scales}a). Interestingly, we find $L_{trans}$ to be, at most, marginally dependent on channel width, while $L_{long}$ clearly increases in wider channels. In the following, we propose scaling arguments for these dependencies, relating them to the relevant material and geometrical parameters and, ultimately, unveiling the physical origin of the emerging spatiotemporal lattices.
 
We observe that $L_{\text{trans}}$ is directly related to $\ell_a$, defined above in the quasi-laminar flow regime. This can be directly visualized in experiments with pillar obstacles  (Fig. \ref{SFig:pillars}), where  $L_{\text{trans}}\simeq 2\ell_a$ in regions where hotspots form. Therefore, we expect $L_{\text{trans}}$ to scale with $\sqrt{K/\zeta}$, as $\ell_a$ does \cite{guillamat2016,Calderer2025}. In our experiments with different channel widths (Fig. \ref{Fig:length_scales}a), we have used the same reference composition of the AN (see Methods), which would suggest that all our preparations feature the same $\zeta$ and $K$ values, and thus the same $L_{\text{trans}}$ for all channels. However, we have observed that the average speed of active flows, $v$, which is a proxy for $\zeta$ in this system \cite{giomi2015,guillamat2016}, decreases when increasing the channel width (Fig. \ref{SFig:velocity_and_thickness}a). This should lead to $L_{\text{trans}}$ increasing with channel width (see Fig. \ref{SFig:spacing} for experimental evidence of $L_{\text{trans}}$ increasing as $v$ decreases). Interestingly, confocal fluorescence microscopy reveals that the AN layer is thicker in narrow channels (see Fig. \ref{SFig:velocity_and_thickness}b), which suggests that $K$ is also larger in narrow channels. Since both $\zeta$ and $K$ have a similar trend, our observation that $L_{trans}$ is roughly independent of channel width becomes plausible.  

On the other hand, the origin of the observed width dependence of $L_{\text{long}}$ needs a more elaborate argument. The starting point is the intrinsic instability of the aligned MTs in the quasi-laminar, unconfined flows \cite{guillamat2016}. Indeed, the aligned extensile filaments in this configuration are prone to suffer bending instabilities that originate transversal flows,\emph{ i.e.}, perpendicular to the alignment direction. An example of the onset and decay of such transient instabilities is shown in Fig. \ref{Fig:transversal}a, along with a sketch of the evolution of AN defects that drive the flows. The synchronization of these instabilities with the quasi-laminar flows is weak in unconfined AN layers, where stable spatiotemporal hotspot lattices are absent. It is, nevertheless, possible to visualize some incipient local patterning when measuring the time average of an order parameter defined for the orientational field (Fig. \ref{Fig:transversal}b). The synchronization between transversal and longitudinal flows is only consolidated under lateral confinement, where the stable spatiotemporal lattice emerges both in the map of the orientation order parameter (Fig. \ref{Fig:transversal}b) and in the density and vorticity fields, as shown above. As discussed, the ability to form and stabilize  this lattice requires high activity. For low activities, transversal instabilities are scarce, and no spatiotemporal lattice can be observed either in the order parameter, density, or vorticity maps, even under lateral confinement (Fig. \ref{Fig:transversal}b; see also Fig. \ref{SFig:cagedATP} and Movie \ref{SMov:cagedATP}).

Transversal instabilities occur at a characteristic rate, related to the intrinsic time scale of the AN, $\tau \sim \eta/\zeta$, at which viscous and active stresses balance \cite{giomi2015,guillamat2016}. Here, $\eta$ is the effective dynamic viscosity of the AN. For quasi-laminar active flows with average longitudinal speed $v_x$, aligned active filaments advance a typical distance $v_x\, \tau$ before the next transversal instability occurs, thus determining a coherence length in the longitudinal direction, which we relate to $L_{\text{long}}$.  In Fig. \ref{Fig:length_scales}b we report the intrinsic oscillation frequency of the director field (Fig. \ref{Fig:setup}e) as a function of the flow speed. From those measurements, we obtain $\tau$ and compare the experimental values of $L_{\text{long}}$ with $v_x \, \tau$ (see inset). Indeed, we find an excellent agreement except from some outlying experiments. We therefore, propose the scaling $L_{\text{long}}\propto v_x \tau \propto v_x \eta/\zeta$.
We have seen that $v$ increases for narrow channels (Fig. \ref{SFig:velocity_and_thickness}a) and $v\propto\zeta$ \cite{guillamat2016}, and similar trends are expected for $v_x$. Moreover, it has also been measured that this AN material is shear thinning, with $\eta$ smaller for larger speeds \cite{velez2024}. As a consequence, we expect $L_{\text{}long} \propto \eta$ and, thus, to decrease for narrower channels where speed is higher, as we indeed observe in experiments (see Fig. \ref{Fig:length_scales}). 

Besides validating the observed trends of the lattice parameters using scaling arguments, the conclusion of this analysis is that the observed spatiotemporal crystals are the result of the synchronization of longitudinal flows, imposed by the interfacial anisotropy on the AN, with transversal flows, intrinsic to the extensile nature of the active filaments. Both types of flows organize in orthogonal sets of antiparallel flow lanes, as illustrated in Fig. \ref{Fig:transversal}c. The combination of these flows organize both the alternated lattice of vortices and a complementary lattice of stagnation regions, where density hotspots organize (see sketch in Fig. \ref{Fig:transversal}d). 


In the next section, we reveal a crosstalk mechanism between the active and passive fluids that are fundamental for the stabilization of the spatiotemporal active crystal under lateral confinement. 

\subsection{Crosstalk between the active and the passive phases leads to active crystallization}
\begin{figure}[t]
\centering
\includegraphics[width=\columnwidth]{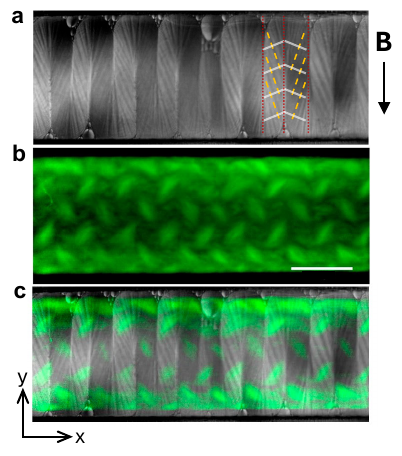}
\caption{Crostalk between active and passive phases. \textbf{a} Polarized microscopy image of the passive liquid crystal layer showing the SmA patterning. In the sketch, dashed lines indicate the orientation of the passive mesogens, solid lines indicate the arrangement of the SmA planes, and dotted lines are the curvature walls forced by geometry on the SmA planes. \textbf{b} Time average of 480 fluorescence micrographs of the AN layer acquired at 2 fps. Scalebar, 200 $\mu$m.  \textbf{c} Images \textbf{a} and \textbf{b} are overlaid to highlight their spatial correlation.}
\label{Fig:spots_feathers}
\end{figure}
We have seen above that confinement stabilizes the intrinsic tendency of the aligned AN to organize a spatiotemporal crystal. In this section, we show that, besides providing with the anisotropic interfacial viscosity that triggers the aligned quasi-laminar flows, the SmA layer mediates in the pattern stabilization through a feedback mechanism. 

As seen in Fig. \ref{SFig:velocity_and_thickness}b, the water-oil interface features a curved meniscus across the channel, since the aqueous phase preferentially wets the polymeric walls (see Methods). Layered mesophases, such as the SmA liquid crystal used here, form undulated structures and buckle when they adapt to frustrated curved geometries, in an attempt to maintain the equilibrium layer spacing. This is known as the Helfrich-Hurault instability \cite{Blanc2023}. Polarizing optical microscopy reveals  patterns along the channels in the SmA phase similar to those reported for spherical shells \cite{Lopez-Leon2012} (Fig. \ref{Fig:spots_feathers}a). The SmA planes, which determine the easy-flow direction of the AN, slightly depart from their orientation perpendicular to the applied magnetic field. Instead, they adopt the indicated zig-zagging structure with curvature walls that accumulate the distortions caused by the curved interface on the lamellar arrangement \cite{Lopez-Leon2011}. 

Remarkably, the longitudinal periodicity of the curvature walls matches exactly $L_{long}$ in the AN density hotspot lattice, as we can see in Fig. \ref{Fig:spots_feathers}b and c (see also Fig. \ref{SFig:fully_developed} and Movie \ref{SMov:transition_feathers}) and in Fig. \ref{SFig:spacing}, where we observe that the spacing of curvature walls follows $L_{long}$ at all activities. 

An effective crosstalk, and thus formation of the spatiotemporal crystal in the AN and the concomitant arrangement of curvature walls in the SmA is only realized at high enough activities (see Movie \ref{SMov:feather_formation}). Indeed, formation of the curved SmA layer in contact with an inactivated AN layer, i.e., with arrested flows, leads to a random arrangement of curvature walls that does not appear to depend on channel width, consistently with the fact that interface curvature does not change significantly with the width of the channel (see Fig. \ref{SFig:velocity_and_thickness}b). 

The presence of this SmA pattern is crucial to stabilize the AN spatiotemporal crystal, as evidenced from experiments where cylindrical polymeric columns are embedded as obstacles in the system. The water-oil interface has a curved meniscus between nearest columns but it is flat elsewhere. This leads to the formation of the feather-like pattern in the same regions where the spatiotemporal AN crystal form (see Fig. \ref{SFig:pillars_feathers}). A final evidence is provided in Fig. \ref{SFig:nocurvature} and Movie \ref{SMov:nocurvature}, where we have used an alternative in situ microfabrication technique to build hydrogel walls \cite{velez2024} so that the confined aqueous/oil interface is flat. As a result, the feather-like pattern is absent (see Fig. \ref{SFig:nofeathers}), and we observe the same quasi-laminar AN flows that we obtain in absence of lateral confinement. 

\begin{figure}[t]
\includegraphics[width=0.9\columnwidth]{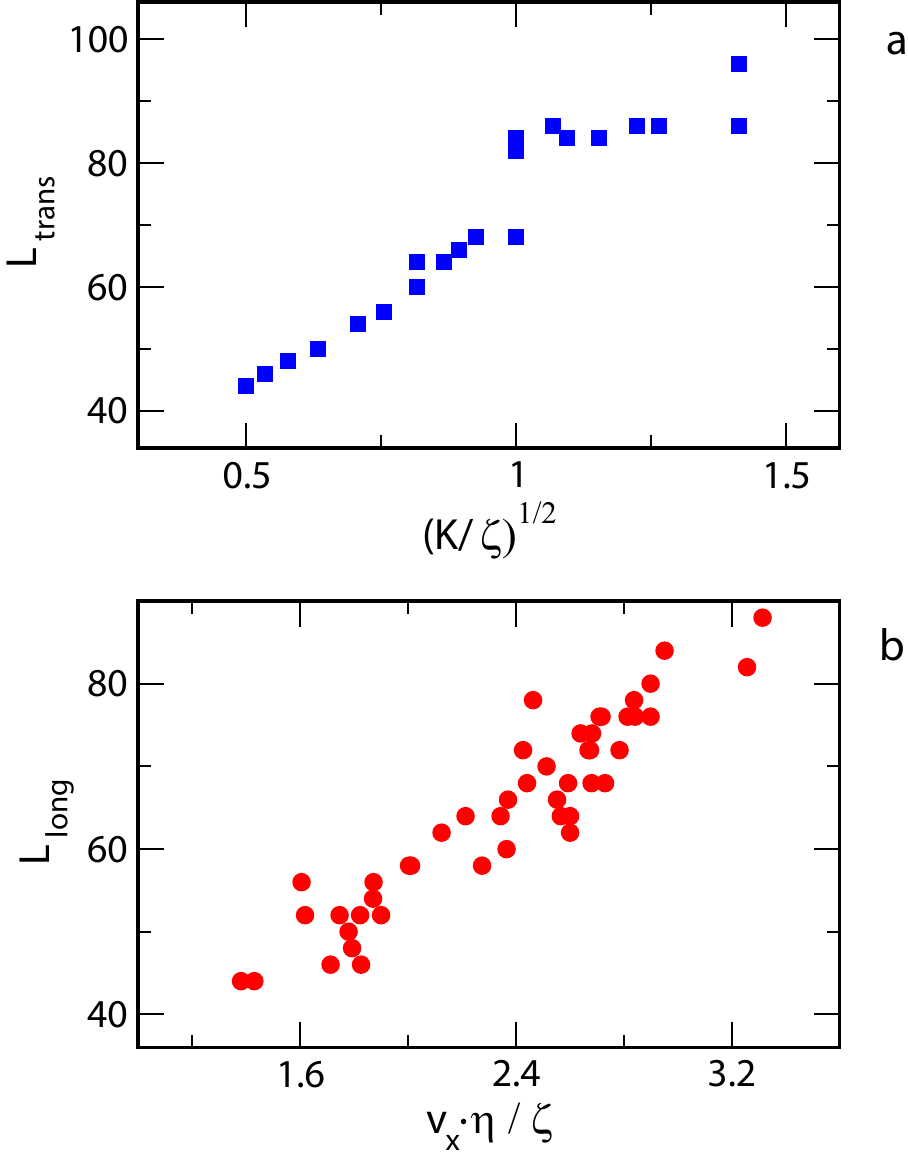}
\caption{Scaling of spatiotemporal patterns in simulations (Model A). \textbf{a} Scaling for $L_{\text{trans}}$ length obtained from nematic simulations using a kinematic viscosity $\eta = 6.67$, and different values for $K \in (0.005-0.05)$ and $\zeta \in (0.005-0.05)$. \textbf{b} Scaling for $L_{\text{long}}$ length obtained from our simulations using a fixed value for $K = 0.015$ and different values for $\eta \in (6.67-22.7)$ and $\zeta \in (0.005-0.02)$. All parameter magnitudes are provided in simulation units although the characteristic spacings shown here can be dimensionalized by considering that a simulation length unit approximately corresponds to $1.25 \, \mu m$ (see Methods).} 
\label{FigScalings}
\end{figure}

\subsection{Emergence of spatiotemporal AN crystals in numerical simulations}

In this section, we present results from a numerical analysis of this system, with a model that reproduces the basic qualitative experimental results and confirms the scaling relations predicted for the lattice parameters. To model the dynamics of the AN layer we use a continuum two-dimensional description of an active gel \cite{ramaswamy10, marchetti2013, Prost15, doostmohammadi2018active}.
No-slip and strong planar anchoring are imposed on the channel walls, and periodic boundary conditions are considered in the direction along the channel. 
The dynamic equations and their hybrid numerical resolution method~\cite{marenduzzo2007steady, C9SM00859D} are summarized in detail in the Materials and Methods section.

In our study, we simulate only the active layer, but introduce a friction tensor $\chi_{\alpha\beta}$, which represents the hydrodynamic coupling with
the underlying anisotropic SmA phase. 
We first adopt a simplified approach (Model A) in which we assume that the SmA planes are exactly aligned along the channel walls. The effect of the SmA on the AN is then modeled by a diagonal $\chi_{\alpha\beta}$, where the friction component in the direction of the channel width, $\chi_{yy}$, is much larger than in the easy flow direction, $\chi_{xx}$. 

This approach reproduces the formation of quasi-laminar antiparallel flow lanes, aligned with the $\it{easy}$ flow direction, which separate bands of high nematic order, as observed experimentally. These structures define the characteristic length scale along the $Y$-axis, $L_{\text{trans}}$. As in experiments, this quasi-laminar state is periodically disrupted by transversal flow instabilities and by the nucleation of topological defects that travel across flow lanes (Movie \ref{SMov:ModelA}). At this stage, a second characteristic length emerges along the longitudinal direction, $L_{\text{long}}$. We characterized the geometry of the emerging spatiotemporal patterns by measuring the spatial correlations of the velocity components, which proved to be a consistent method to reveal $L_{\text{trans}}$ and $L_{\text{long}}$ (Fig. \ref{SFig:vcorr}a,b). We have verified that the same procedure can also be consistently applied in experiments (Fig. \ref{SFig:vcorr}c,d). See Materials and Methods for a detailed description of this analysis. We perform simulations with different values of $K$ and $\zeta$, confirming the scaling $L_{\text{trans}} \propto \sqrt{K/\zeta}$ and $L_{\text{long}}\propto v_x \eta / \zeta$ (Fig. \ref{FigScalings}), thus validating the conjectures raised earlier to explain the dependence of the spatiotemporal lattice in experiments. 

\begin{figure}[t]
\includegraphics[width=\columnwidth]{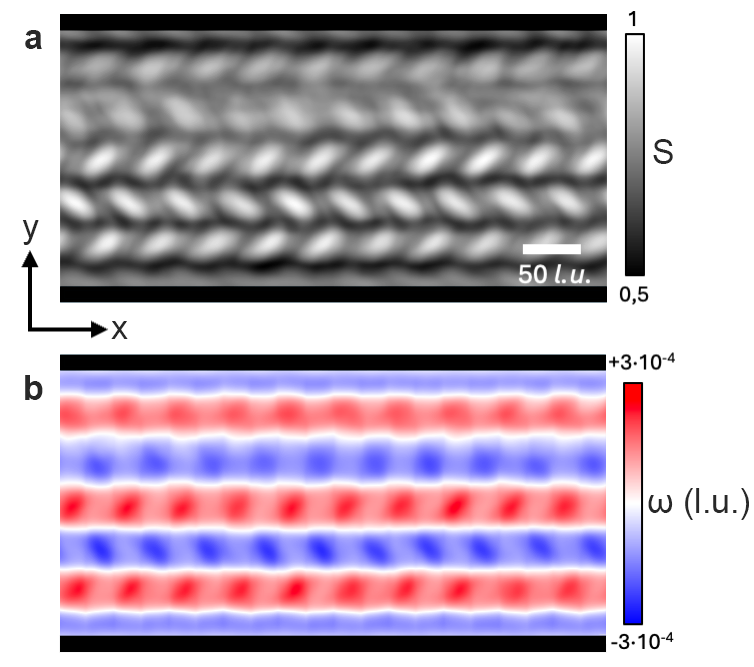}
\caption{Time averaged patterns of the nematic scalar order parameter (\textbf{a}) and vorticity (\textbf{b}), for the reference case in Model B (friction tensor with fixed vertical stripes of alternating nondiagonal friction terms, $\chi_{xy}= \pm 0.01$). The spatial periodicity of the torque stripes is $512/10$, that approximates the longitudinal spacing $L_{long} \approx 50\,l.u.$ obtained from Model A (with a diagonal friction tensor,\textit{ i.e.}, $\chi_{xy}= 0$). The two characteristic spacings are consolidated and the pattern of hotspots and the antiferromagnetic vortex lattice are clearly observed.}
\label{FigHotSpots}
\end{figure}

We next address the proposed role in the stabilization of the emergent spatiotemporal lattice of the transversal SmA curvature walls, which we demonstrated in experiments to adopt a zig-zagging pattern with a periodicity given by $L_{\text{long}}$  (see Fig. \ref{Fig:spots_feathers}). To include this arrangement in simulations, we assume that the off-diagonal components of $\chi_{\alpha\beta}$ are non-zero, and instead alternate between two values $\pm \chi_{xy}^{(0)}$ along the channel, with a periodicity given by $L_{\text{long}}$. We will refer to this approach as Model B. For example, in the reference case analyzed in Materials and Methods with Model A ($\eta=6.67$, $K=0.015$, $\zeta=0.0075$, $\chi_{xx}=0.004$ and $\chi_{yy}=0.1$), we obtained a longitudinal spacing $L_{long} \approx 50\,l.u.$ (Fig. \ref{SFig:vcorr}b). We use the same parameter to simulate Model B, where the periodicity of $\chi_{xy}= \pm 0.02$ is set to approximately $L_{long}$. The results of this simulation are shown in Fig. \ref{FigHotSpots} and Movie \ref{SMov:ModelB}, where the emergence of a herringbone lattice in the form of hotspots of the nematic order parameter and alternating vortices is clearly visible, consistent with experiments. \\

\section{Discussion}

We have demonstrated that order and chaos can be reconciled in an active fluid where a crystalline lattice of density, orientation, and vorticity emerges when self-sustained chaotic flows adapt to suitable boundary condition. This is realized through the interplay between activity, anisotropy and topology in a system where a self-propelled material is interfaced with a passive liquid crystal oil. Indeed, the reported phenomena arise through a subtle feedback between the microtubule/kinesin active nematic layer and the hydrodynamically coupled SmA mesophase. 

The active layer drives the soft matter system, while the viscous anisotropy of the passive phase induces reorganization into quasi-laminar active flows. In its turn, aligned extensile active filaments are intrinsically unstable, promoting topological defect unbinding that drives random transversal flows. On the other hand, the layered structure of the passive phase has to adapt to the topological requirements of the curved water/oil interface within microfluidic channels. As a result, SmA planes buckle into a zig-zagging pattern along the channels through the Helfrich-Hurault instability. This perturbs the otherwise longitudinal easy flow direction of the AN. Remarkably, the scale of the zig-zagging pattern is defined by the intrinsic time scale of the active flows that, this way,  synchronizes transversal and longitudinal flows.  

The two spatial scales that characterize the emerging active patterns have disparate origins: while the transversal periodicity encodes the active length scale, which is a balance between elastic and active stress, the longitudinal periodicity is the result of the balance between active injection and viscous dissipation, which encodes an intrinsic time scale. We support this interpretation with numerical simulations using an active nematohydrodynamic model where coupling with the underlying SmA phase is approximated as a friction stress. This numerical framework will be the starting point of an eventual first-principles study of the hydrodynamical coupling between the active and the passive anisotropic fluid layers.


We have shown that the symmetry of active turbulent flows can be  broken both in space, through the spontaneous formation of regular patterns of density, orientation, and vorticity fields, and in time, through the emergence of periodic oscillations of the local fields. This behavior is reminiscent of continuous time crystals, \textit{i.e.}, non-equilibrium states of matter that break the time-translation symmetry by settling into periodic oscillations while receiving a continuous, non-periodic input of energy \cite{Sacha2018,Kongkhambut2022,Zalatel2023}. Just recently, a continuous time crystal has been realized outside of the quantum and photonic realm, in an excited nematic liquid crystal device \cite{Zhao2025}. In our active system, internally stored energy is harnessed into a spatiotemporal crystal, whose spatial and temporal periodicity are dictated by the material parameters. Unlike all previous examples, ours is a closed system, resulting in a more faithful realization of the original concept of time crystal \cite{Shapere2012}. Crystallizing active turbulence by leveraging the active dynamics and the intrinsic length and time scale of an active nematic system is a generic strategy that can be spontaneously adopted or imposed through suitable boundary conditions, allowing to envision new biotechnological applications. 


\section{Materials and methods}\label{sec:methods}

\subsection{Thermotropic Liquid Crystal }

4-cyano-4$^{'}$-octylbiphenil (8CB) (SYNTHON Chemicals GmbH $\&$ Co.; ST01422) is a thermotropic liquid crystal between 21.5 and 40.4 °C, featuring a lamellar Smectic-A (SmA) phase in the temperature range 21.5 $^{\circ}$C $< T <$ 33.5 $^{\circ}$C. 8CB is diamagnetic and exhibits positive diamagnetic anisotropy ($ \chi_{a} \sim {10^{-6}}$ ). 

\subsection{Polymerization of MTs}

Stabilized microtubules (MTs) were polymerized from heterodimeric ($\alpha, \beta$)-tubulin from bovine brain (Biomaterials Facility, Brandeis University MRSEC, Waltham, MA) with a non-hydrolyzable GTP analog, guanosine-5-[($\alpha, \beta$)-methyleno]triphosphate (GMPCPP) (Jena Biosciences; NU-405), which suppresses tubulin turnover \cite{Hyman1992}. Tubulin labeled with Alexa 647 was mixed with unlabeled fraction of the protein till final concentration of the label 3 ${\%}$ and, incubated with 0.6 mM GMPCPP and 1.2 mM ditiotreitol (DTT) (Sigma-Aldrich; 43815) in M2B buffer (80 mM Pipes, 1 mM EGTA, 2 mM MgCl2, pH 6.8; Sigma-Aldrich; P1851, E3889, M4880, respectively) first at 37 $^{\circ}$C for 30 min and after for 4h at room temperature. The last step determines the MTs length distribution, which is typically $ \sim {1-2} \mu m $. 

\subsection{Kinesin expression}

Construct encoding \textit{Drosophila melanogaster} heavy-chain kinesin-1 fragment (K401-BIO-H6: 401 aa of the N-terminal kinesin-1 fused to 87 aa of C-terminal of the biotin carboxyl carrier protein subunit of \textit{E. coli} and His6-tag) (addgene id: 15960) was expressed in \textit{E. coli} Rosetta (DE3) and purified with affinity (HisTrap\textsuperscript{TM} Fast Flow column) (Sigma-Aldrich; GE17-5255-01) and size-exclusion chromatography (Superdex\textsuperscript{\textregistered} 75 10/300 GL) (Sigma-Aldrich; GE17-5174-01) using FPLC (AKTA go\textsuperscript{TM} chromatography system). The purified protein was concentrated with the kinesin concentration column (Millipore; Amicon\textsuperscript{\textregistered} Ultra 10K) and its concentration was estimated on the SDS PAGE gel by the intensity of the bands versus calibration with the BSA protein standards (Bovine serum albumin) (Sigma-Aldrich; P0834). The protein was stored in a 30${\%}$ (w/v) aqueous sucrose solution at -80 $^{\circ}$C for future use. 

\subsection{Assembly of the active gel }

The kinesin motor clusters were prepared by mixing biotinylated kinesin motor proteins and tetrameric streptavidin (Invitrogen; 434301) in molar ratio 2:1 in the presence of 0.22 mM DTT. The mixture obtained was incubated on ice for 30 min. 

The dimerized motor units were then mixed with the feeding solution, containing ATP (Sigma; A2383), the ATP-regenerating system (phosphoenolpyruvate (PEP) (Sigma; P7127), pyruvate kinase/lactate dehydrogenase (PK/LDH) (Sigma; P0294), a nonadsorbing polymeric agent (PEG, 20 kDa; Sigma; 95172)  that promoted the formation of filament bundles through depletion, oxygen scavenging system and antioxidants (Catalase (Sigma-Aldrich; C40),  Glucose Oxidase (Sigma-Aldrich; G2133), D-(+)-Glucose (Sigma-Aldrich; G7021), 6-Hydroxy-2,5,7,8-tetramethylchromane-2-carboxylic acid (Trolox) (Sigma-Aldrich; 238813), DTT (Sigma-Aldrich; 43815)) and the PEG-based triblock copolymer surfactant Pluronic F-127 (Sigma; P-2443). MTs were added to the resulting mixture just before starting the experiment. The final concentration of all ingredients is listed in Table 1.
\begin{table}[h]
    \centering
    \caption{Composition of active material. \textsuperscript{*}M2B buffer: 80 mM PIPES pH 6.8, 2 mM MgCl\textsubscript{2}, 1 mM EGTA. \textsuperscript{**}Phosphate buffer: 6.68 mM KH\textsubscript{2}PO\textsubscript{4}, 12.32 mM K\textsubscript{2}HPO\textsubscript{4} pH 7.2.}
    \label{tab:placeholder_label}
    \begin{tabular}{p{0.45\columnwidth} p{0.45\columnwidth}}
        \toprule
        \textbf{Compound}      & \textbf{Final concentration} \\ 
        \midrule
        PEG (20 kDa)\textsuperscript{*}        & 1.54\% w/v      \\
        PEP\textsuperscript{*}                 & 25.68 mM        \\
        MgCl\textsubscript{2}\textsuperscript{*} & 3.12 mM       \\
        Trolox\textsuperscript{**}             & 1.93 mM         \\
        ATP\textsuperscript{*}                 & 1.37 mM         \\
        Catalase\textsuperscript{**}           & 0.04 mg/ml      \\
        Glucose\textsuperscript{**}            & 3.20 mg/ml      \\
        Glucose Oxidase\textsuperscript{**}    & 0.21 mg/ml      \\
        PK/LDH                                 & 25.01 / 24.91 U/ml \\
        Pluronic-147\textsuperscript{*}        & 0.41\% w/v      \\
        DTT\textsuperscript{*}                 & 5.21 mM         \\
        Streptavidin\textsuperscript{*}        & 0.01 mg/ml      \\
        Kinesin\textsuperscript{*}             & 0.08 mg/ml      \\
        Microtubules\textsuperscript{*}        & 1.85 mg/ml      \\
        \bottomrule
    \end{tabular} 
\end{table}

\subsection{Active nematic cell}

The AN is formed at the interface with anisotropic oil, 8CB, where Pluronic-147 surfactant ensures a planar alignment of 8CB molecules at the water/LC interface. Active flows were studied in two different geometries: open and closed cells.  

For the open cell experiments, PDMS-pools with four open channels or arrays of circular pillars on the bottom were fabricated by casting poly-dimethylsiloxane (Sylgard\textsuperscript{\textregistered}, Dow Corning) using a custom mold (see Fig.1). PDMS-pools had diameter of 6 mm and depth of 1 mm and just before experiment were pretreated with oxygen plasma followed by 2${\%}$ Pluronic F-127 solution for 30 min at room temperature, to make surface hydrophilic and increase the life lasting of AN. The PDMS block was glued onto round cover glass of 18 mm diameter (United Scientific Supplies; USS-SQCVRSP18). Active gel mix was loaded on the bottom of the pool, and it filled channels or interpillar space with capillarity. After the active drop was evenly distributed along the PDMS pool, it was immediately covered with 8CB LC preheated till isotropic state. 

The closed cell experiments were carried out using flow cells with a channel of 1.5-2 mm width and 150 $\mu$m height. The cell was assembled from superhydrophilic polyacrylamide-coated glass and hydrophobic one treated with dimethyl(octadecyl)[3-(trimethoxysilyl)propyl]ammonium chloride (DMOAP) (Thermo Scientific, 338531000), separated by a 150 $\mu$m thick double-sided tape. The cell was filled by capillarity first with 8CB LC, followed by introducing the active material. To avoid evaporation, the cell was sealed using petroleum jelly. To build channels without meniscus inside flow cell, the constituents of the photopolymerizing hydrogel were incorporated in solid form to the active mixture: photoinitiator lithium phenyl-2,4,6-trimethylbenzoylphosphinate (LAP)  (TCI; L0290) and the monomer 4-Arm PEG-Acrylate 5 kDa (4PEG5k) (Biochempeg; A44009-5k) till final concentration 0.226 and 4.52 ${\%}$ w/v correspondingly. The photopolimerisation of hydrogel was performed with UV light projector as described earlier \cite{velez2024}. 

\subsection{Experimental Setup}

The experimental setup was previously described \cite{Bantysh2024}. Briefly, samples were enclosed by a custom cylindrical thermostatic oven and placed in the 25-mm-wide cylindrical cavity of a 0.4T Halbach array formed by eight permanent magnets (Bunting Magnetics Europe Ltd.), where the magnetic field is homogeneous. The system was heated up to 34 $^{\circ}$C to promote SmA to N transition of 8CB and incubated at this temperature up to 30 min, till the active material in the gel spontaneously condensed onto the water/8CB interface, leading to the formation of the active nematic layer. After that, temperature was reduced to 33.00 $^{\circ}$C, leading N to SmA transition of LC and corresponding alignment transition of AN flows. 

\subsection{Imaging}

The routine observation of the sample was carried out with a custom-made microscope allowing observation in fluorescent or polarized mode. Epifluorescence of MTs was excited with a white LED source (Thorlabs MCWHLP1) and a Cy5 filter set (Edmund Optics) and captured with Iris9 CMOS camera (Teledyne Photometrics) operated with ImageJ $\mu$-Manager open-source software. In polarized mode a low voltage halogen lamp (30W, 6V) (Philips; 5761) was used as a source of light. For closed cell experiments, a Nikon Eclipse Ti2-e equipped with a white LED source (Thorlabs MWWHLP1) and a Cy5 filter cube (Edmund Optics) was used, also operated with ImageJ $\mu$-Manager. Confocal acquisition in fluorescence mode was performed with Zeiss LSM880 high speed spectral confocal microscope. 

\subsection{Image analysis}

Experimental images were preprocessed with ImageJ and prepared for followed up extraction and analysis of director or velocity fields. The first one was extracted with a customized MATLAB code based on \cite{Ellis2020}, the latter one using a PIV application implemented in MATLAB \cite{Thielicke2021}. Director-based order parameter was calculated as described in \cite{Bantysh2024}. 

\subsection{Active nematic simulations}

\subsubsection{Model, method and equations}
We simulate the active nematic using a two-dimensional continuum hydrodynamic framework.
The local orientation of the filaments is captured by the director field $\mathbf{n}(\mathbf{x})$, while the
scalar order parameter $S$ quantifies the degree of orientational order, with $S=0$ corresponding to the isotropic
phase and a finite value $S_N$ representing nematic phase.
Due to the apolar nature of the filaments (head-tail symmetry), we employ a tensorial order parameter defined
as $Q_{\alpha \beta} = 2S (n_\alpha n_\beta - \delta_{\alpha \beta}/2)$.

In this model, the bulk contribution to the free energy is omitted. This choice reflects experimental observations
in microtubule-kinesin
suspensions, where nematic order arises solely due to activity. Similar modeling approaches have been employed in prior
studies~\cite{Thampi_2015, Hardouin2019,  velezceron2025activenematicpumps}.
The elastic contribution to the free energy penalizes spatial distortions in the order parameter and is given by
\[
f_E = \frac{K}{2}(\partial_\gamma Q_{\alpha \beta})^2
\]
where $K$ is the elastic constant, assuming a one-elastic constant approximation. The total free energy corresponds to the integration over the entire area of the nematic layer
\[
\mathcal{F} = \int f_E\, d^2x.
\]

The system dynamics follow the Beris–Edwards formalism for the evolution of the order parameter tensor $Q_{\alpha\beta}$,
coupled to the continuity and Navier–Stokes equations governing the velocity field $\mathbf{v}(\mathbf{x})$:
\begin{align}
  &\partial_t Q_{\alpha \beta} + v_\gamma \partial_\gamma Q_{\alpha \beta} - S_{\alpha \beta} = \Gamma H_{\alpha \beta}, \label{eq:beris-edwards} \\
  &\partial_\alpha v_\alpha = 0, \label{eq:continuity} \\
  &\rho \left(\partial_t v_\alpha + v_\beta \partial_\beta v_\alpha\right) = -\partial_\alpha p + 2\eta \partial_\beta D_{\alpha \beta} \nonumber\\& - \zeta \partial_\beta Q_{\alpha \beta} - \chi_{\alpha\beta}v_\beta. \label{eq:navier-stokes}
\end{align}
Here $\Gamma$ stands for the rotational diffusivity, $H_{\alpha \beta}$ is the molecular field that reads
\[
H_{\alpha \beta} = -\frac{\delta \mathcal{F}}{\delta Q_{\alpha \beta}} + \frac{\delta_{\alpha \beta}}{2} \mathrm{Tr} \left(\frac{\delta \mathcal{F}}{\delta Q_{\gamma \epsilon}} \right),
\]
and $\chi_{\alpha\beta}$ denotes a tensor describing the friction between the active nematic layer and the SmA subphase.

The co-rotational term $S_{\alpha \beta}$ accounts for the coupling between flow and orientation and is given by
\begin{eqnarray}
\nonumber
S_{\alpha \beta} = (\xi D_{\alpha \gamma} + W_{\alpha \gamma})(Q_{\beta \gamma} + \delta_{\beta \gamma}/2) + (Q_{\alpha \gamma} + \delta_{\alpha \gamma}/2)\\
\nonumber
\times(\xi D_{\gamma \beta} - W_{\gamma \beta}) - 2\xi(Q_{\alpha \beta} + \delta_{\alpha \beta}/2)(Q_{\gamma \epsilon} \partial_\gamma v_\epsilon),
\end{eqnarray} 
where the strain rate and vorticity tensors are defined as
$D_{\alpha \beta} = (\partial_\beta v_\alpha + \partial_\alpha v_\beta)/2$
and $W_{\alpha \beta} = (\partial_\beta v_\alpha - \partial_\alpha v_\beta)/2$, respectively.

The alignment parameter $\xi$, which
reflects the shape of the constituent particles, is positive for rod-like filaments and negative for disk-like ones.
We assume rod-shaped filaments in the flow-aligning regime. In the momentum equation, $\rho$ is the fluid density, $p$ is the pressure,
$\eta$ is the shear viscosity (the kinematic viscosity $\nu = \eta / \rho$), and $\zeta$ quantifies the strength of activity. Since active stresses dominate in this regime~\cite{velez2024},
we neglect the passive elastic stress term in Eq.~\eqref{eq:navier-stokes}, following the approach of prior
studies~\cite{Hardouin2019, PhysRevE.106.014705}.

For the numerical implementation, Eq.~\eqref{eq:beris-edwards} is solved using a predictor-corrector finite difference method,
while Eqs.~\eqref{eq:continuity} and~\eqref{eq:navier-stokes} are recovered using a lattice Boltzmann
algorithm~\cite{marenduzzo2007steady, C9SM00859D}. At solid boundaries, we impose strong planar anchoring and no-slip conditions
using a halfway bounce-back scheme, which reflects experimentally realistic interactions. 

To mimic the experimental channels, simulations are run in systems of size $N_x \times N_y$, with periodic boundary conditions in the $X$-direction and solid walls at $y = 0$ and $y = Ny + 1$. As in experiments, we consider the SmA planes oriented along the $X$-axis so that the \textit{easy} direction is parallel to the solid walls 
($\chi_{yy} \gg \chi_{xx}$).
Initial conditions correspond to a quiescent fluid with directors mostly aligned along the horizontal axis, with random angular
deviations of $\pm 2^\circ$. Different random seeds are used across simulations. Unless stated otherwise, we simulate systems with $N_x=512$ and $N_y=256$ and the following parameters
are used in simulation units: alignment parameter $\xi = 0.9$, density $\rho = 40$,
rotational diffusivity $\Gamma = 0.4$, $\chi_{yy}=0.1$ and $\chi_{xx}=0.004$. The other parameters are given in the text and the corresponding figure captions.

\subsection{Obtaining pattern length scales from active nematic simulations}

Numerical simulations using a diagonal anisotropic friction tensor $\chi_{\alpha\beta}$ reveal that an active nematic layer in contact with an SmA subphase organizes into bands
along the \textit{easy} direction with alternating velocity directions. In Movie \ref{SMov:ModelA} we show the temporal evolution for the reference case corresponding to $\eta=6.67$, $K=0.015$, $\zeta=0.0075$, $\chi_{yy}=0.1$ and $\chi_{xx}=0.004$. The two characteristic length scales arising from this scenario can be readily obtained
from our simulations by analyzing appropriate autocorrelation functions (Fig. \ref{SFig:vcorr}a,b).
In particular, the band width is extracted from the spatial autocorrelation function in the Y-coordinate of the X-component of
the local nematic velocity field:
\[
C_y (y)= \left< v_x(x_0,y_0) \cdot v_x(x_0,y_0+y) \right>
\]
where the brackets denote an average taken first over all lattice sites $x_0,y_0$, and then over time across different temporal frames. The distance of the first maxima of the autocorrelation corresponds to the spatial periodicity of the formed bands
(for our reference case, $L_{trans} \approx 86 \, l.u.$, see Fig. \ref{SFig:vcorr}a). 

The characteristic length scale associated with the transverse instabilities that arise between bands of oppositely directed flow jets is also computed
using spatial autocorrelation functions. In this case, the spatial correlation is calculated along the X-axis for the Y-component of the local velocity field,
\[
C_x (x)= \left< v_y(x_0,y_0) \cdot v_y(x_0+x,y_0) \right>
\]
where the brackets denote average over all lattice sites $x_0,y_0$ at different temporal frames. It is important to note that these instabilities emerge spontaneously at different locations along the various jet flows.
Although they may transiently synchronize across the width of the system,
they ultimately vanish upon temporal averaging of the nematic system fields.
For this reason, when estimating this characteristic length scale, it is crucial to first compute the autocorrelation function at each individual time frame
and then average the resulting correlation functions. The distance of the first maxima of this autocorrelation corresponds to the spatial periodicity of the
spontaneous transversal fluctuations (for our reference case, $L_{long} \approx 50 \, l.u.$, see Fig. \ref{SFig:vcorr}b).

\subsection{Dimensional analysis of simulation magnitudes}

The parameters used in the simulations are expressed in simulation length, time, and mass units. Specifically, the lattice spacing is set to $\Delta x = 1$, the time step to $\Delta t = 1$, and the reference density to $\rho_{r} = 1$, thereby defining the system of simulation units. All variables and parameters are reported within this framework, although a correspondence with physical dimensions can be established by comparing representative experimental properties with their simulated counterparts.

For the specific case of length units, for example, the structure of an unconstrained active nematic is characterized by an intrinsic active length scale, defined by the average defect spacing, $d$ \cite{giomi2015}, which can be readily measured in both experiments and simulations. For a free active nematic (i.e., without contact with any smectic subphase) under the experimental conditions used in this study, we measured $d \approx 70 \, \mu m$ \cite{velez2024}; whereas for the corresponding simulations of a free active nematic ($\chi_{\alpha\beta}=0$) using the parameter set described here, we obtained $d \approx 56 \, \Delta x$ \cite{velez2024}. This comparison yields an estimate of the physical size of $\Delta x \approx 1.25 \, \mu m$, indicating that the characteristic spacings obtained from experiments and simulations are closely comparable and of the same order of magnitude (compare Figs. \ref{Fig:length_scales} and \ref{FigScalings}), while the dimensions of the simulated systems are likewise consistent with those observed experimentally.

In order to verify that the simulations were carried out under conditions consistent with the experimental ones, we can, for instance, compare the values of the activity used in both contexts. To this end, it is necessary to consider not only the units of length, but also those of time and mass. Since activity has units of shear viscosity ($\eta = \nu \rho$) times velocity per length, we can compute a sort of reference activity $\zeta_{ref} = \nu \rho v_{r}/d$, where $v_r$ is the mean velocity of the active nematic. In our simulations for a free active nematic, $v_{r}=0.0105 \,\Delta x/\Delta t$ in simulation units. Taking the values of density and viscosity used in our simulations, $\rho=40\, \rho_r$ and $\nu=0.167 \, \Delta x^2/\Delta t$, we obtain a reduced activity $\zeta/\zeta_{ref} \approx 6$ for the reference case shown in Fig. \ref{FigHotSpots}. The same estimation can be done for the experiments using $\zeta \approx 2 \cdot 10^{-7} \, Pa \cdot m$ and $\eta =\nu\rho \approx 4 \cdot 10^{-6} \, Pa \cdot s \cdot m$, $d \approx 70 \, \mu m$ and $v_r \approx \, 1 \mu m \cdot s^{-1}$ \cite{velez2024}, yielding a similar value for the reduced activity $\zeta/\zeta_{ref} \approx 4$, thus ensuring that simulations are performed in conditions that match the experimental ones.

\vspace{1cm}

\textbf{Acknowledgements}

The authors are indebted to the Brandeis University MRSEC Biosynthesis facility for providing the tubulin. We thank M. Pons and A. Fernández (Universitat de Barcelona) for their assistance in the expression of motor proteins. We also thank the Scientific and Technological Centers (CCiTUB), Universitat de Barcelona, for their support and advice on the confocal microscopy measurements. O. B., J.I.-M., R. R., and F.S. acknowledge funding from MICINN/AEI/10.13039/501100011033 (Grant No. PID2022-137713NB-C21). O. B. acknowledges a Joan Oró FI fellowship (2023 FI-3 00065) from Generalitat de Cartalunya and the European Social Plus Fund. Brandeis University MRSEC Biosynthesis facility is supported by NSF MRSEC 2011846. R.C. acknowledges financial support from the Portuguese Foundation for Science and Technology (FCT) under the contracts:  2023.10412.CPCA.A2 (DOI: 10.54499/2023.10412.CPCA.A2) and FCT/Mobility/1348751812/2024-25.


\begin{thebibliography}{10}
\expandafter\ifx\csname url\endcsname\relax
  \def\url#1{\burl{#1}}\fi
\expandafter\ifx\csname urlprefix\endcsname\relax\def\urlprefix{URL }\fi
\providecommand{\bibinfo}[2]{#2}
\providecommand{\eprint}[2][]{\url{#2}}
\providecommand{\doi}[1]{\url{https://doi.org/#1}}
\bibcommenthead

\bibitem{Strogatz2003}
\bibinfo{author}{Strogatz, S.}
\newblock \emph{\bibinfo{title}{Sync: The Emerging Science of Spontaneous
  Order}} \bibinfo{edition}{1st} edn (\bibinfo{publisher}{Theia},
  \bibinfo{address}{New York}, \bibinfo{year}{2003}).

\bibitem{Wensink2012}
\bibinfo{author}{Wensink, H.~H.} \emph{et~al.}
\newblock \bibinfo{title}{Meso-scale turbulence in living fluids}.
\newblock \emph{\bibinfo{journal}{Proc Natl Acad Sci U S A}}
  \textbf{\bibinfo{volume}{109}}, \bibinfo{pages}{14308--13}
  (\bibinfo{year}{2012}).

\bibitem{giomi2015}
\bibinfo{author}{Giomi, L.}
\newblock \bibinfo{title}{Geometry and topology of turbulence in active
  nematics}.
\newblock \emph{\bibinfo{journal}{Physical Review X}}
  \textbf{\bibinfo{volume}{5}}, \bibinfo{pages}{031003} (\bibinfo{year}{2015}).

\bibitem{Alert2022}
\bibinfo{author}{Alert, R.}, \bibinfo{author}{Casademunt, J.} \&
  \bibinfo{author}{Joanny, J.-F.}
\newblock \bibinfo{title}{Active turbulence}.
\newblock \emph{\bibinfo{journal}{Annual Review of Condensed Matter Physics}}
  \textbf{\bibinfo{volume}{13}}, \bibinfo{pages}{143--170}
  (\bibinfo{year}{2022}).

\bibitem{Urzay2017}
\bibinfo{author}{Urzay, J.}, \bibinfo{author}{Doostmohammadi, A.} \&
  \bibinfo{author}{Yeomans, J.~M.}
\newblock \bibinfo{title}{Multi-scale statistics of turbulence motorized by
  active matter}.
\newblock \emph{\bibinfo{journal}{Journal of Fluid Mechanics}}
  \textbf{\bibinfo{volume}{822}}, \bibinfo{pages}{762--773}
  (\bibinfo{year}{2017}).

\bibitem{Martinez-Prat2021}
\bibinfo{author}{Martínez-Prat, B.} \emph{et~al.}
\newblock \bibinfo{title}{Scaling regimes of active turbulence with external
  dissipation}.
\newblock \emph{\bibinfo{journal}{Physical Review X}}
  \textbf{\bibinfo{volume}{11}}, \bibinfo{pages}{031065(1--16)}
  (\bibinfo{year}{2021}).

\bibitem{Tan2019}
\bibinfo{author}{Tan, A.~J.} \emph{et~al.}
\newblock \bibinfo{title}{Topological chaos in active nematics}.
\newblock \emph{\bibinfo{journal}{Nature Physics}}
  \textbf{\bibinfo{volume}{15}}, \bibinfo{pages}{1033--1039}
  (\bibinfo{year}{2019}).

\bibitem{Cross1993}
\bibinfo{author}{Cross, M.~C.} \& \bibinfo{author}{Hohenberg, P.~C.}
\newblock \bibinfo{title}{Pattern formation outside of equilibrium}.
\newblock \emph{\bibinfo{journal}{Reviews of Modern Physics}}
  \textbf{\bibinfo{volume}{65}}, \bibinfo{pages}{851--1112}
  (\bibinfo{year}{1993}).

\bibitem{doostmohammadi2018active}
\bibinfo{author}{Doostmohammadi, A.}, \bibinfo{author}{Ign{\'e}s-Mullol, J.},
  \bibinfo{author}{Yeomans, J.~M.} \& \bibinfo{author}{Sagu{\'e}s, F.}
\newblock \bibinfo{title}{Active nematics}.
\newblock \emph{\bibinfo{journal}{Nature communications}}
  \textbf{\bibinfo{volume}{9}}, \bibinfo{pages}{3246} (\bibinfo{year}{2018}).

\bibitem{Sanchez2012}
\bibinfo{author}{Sanchez, T.}, \bibinfo{author}{Chen, D.~T.},
  \bibinfo{author}{DeCamp, S.~J.}, \bibinfo{author}{Heymann, M.} \&
  \bibinfo{author}{Dogic, Z.}
\newblock \bibinfo{title}{Spontaneous motion in hierarchically assembled active
  matter}.
\newblock \emph{\bibinfo{journal}{Nature}} \textbf{\bibinfo{volume}{491}},
  \bibinfo{pages}{431--4} (\bibinfo{year}{2012}).

\bibitem{Shendruk2017}
\bibinfo{author}{Shendruk, T.~N.}, \bibinfo{author}{Doostmohammadi, A.},
  \bibinfo{author}{Thijssen, K.} \& \bibinfo{author}{Yeomans, J.~M.}
\newblock \bibinfo{title}{Dancing disclinations in confined active nematics}.
\newblock \emph{\bibinfo{journal}{Soft Matter}} \textbf{\bibinfo{volume}{13}},
  \bibinfo{pages}{3853--3862} (\bibinfo{year}{2017}).

\bibitem{Hardouin2019}
\bibinfo{author}{Hardo\"{u}in, J.} \emph{et~al.}
\newblock \bibinfo{title}{Reconfigurable flows and defect landscape of confined
  active nematics}.
\newblock \emph{\bibinfo{journal}{Communications Physics}}
  \textbf{\bibinfo{volume}{2}} (\bibinfo{year}{2019}).
\newblock \urlprefix\url{http://dx.doi.org/10.1038/s42005-019-0221-x}.

\bibitem{opathalage19}
\bibinfo{author}{Opathalage, A.} \emph{et~al.}
\newblock \bibinfo{title}{Self-organized dynamics and the transition to
  turbulence of confined active nematics}.
\newblock \emph{\bibinfo{journal}{Proc. Natl. Acad. Sci. U S A}}
  \textbf{\bibinfo{volume}{116}}, \bibinfo{pages}{4788--4797}
  (\bibinfo{year}{2019}).

\bibitem{guillamat2016}
\bibinfo{author}{Guillamat, P.}, \bibinfo{author}{Ignes-Mullol, J.} \&
  \bibinfo{author}{Sagues, F.}
\newblock \bibinfo{title}{Control of active liquid crystals with a magnetic
  field}.
\newblock \emph{\bibinfo{journal}{Proc. Natl. Acad. Sci. U S A}}
  \textbf{\bibinfo{volume}{113}}, \bibinfo{pages}{5498--502}
  (\bibinfo{year}{2016}).

\bibitem{Bantysh2024}
\bibinfo{author}{Bantysh, O.} \emph{et~al.}
\newblock \bibinfo{title}{First order alignment transition in an interfaced
  active nematic fluid}.
\newblock \emph{\bibinfo{journal}{Phys Rev Lett}}
  \textbf{\bibinfo{volume}{132}}, \bibinfo{pages}{228302}
  (\bibinfo{year}{2024}).

\bibitem{Doostmohammadi2016}
\bibinfo{author}{Doostmohammadi, A.}, \bibinfo{author}{Adamer, M.~F.},
  \bibinfo{author}{Thampi, S.~P.} \& \bibinfo{author}{Yeomans, J.~M.}
\newblock \bibinfo{title}{Stabilization of active matter by flow-vortex
  lattices and defect ordering}.
\newblock \emph{\bibinfo{journal}{Nat Commun}} \textbf{\bibinfo{volume}{7}},
  \bibinfo{pages}{10557} (\bibinfo{year}{2016}).

\bibitem{Kongkhambut2022}
\bibinfo{author}{Kongkhambut, P.} \emph{et~al.}
\newblock \bibinfo{title}{Observation of a continuous time crystal}.
\newblock \emph{\bibinfo{journal}{Science}} \textbf{\bibinfo{volume}{377}},
  \bibinfo{pages}{670--673} (\bibinfo{year}{2022}).

\bibitem{Zhao2025}
\bibinfo{author}{Zhao, H.} \& \bibinfo{author}{Smalyukh, I.}
\newblock \bibinfo{title}{Space-time crystals from particle-like topological
  solitons}.
\newblock \emph{\bibinfo{journal}{Nat Mater}} \textbf{\bibinfo{volume}{24}}
  (\bibinfo{year}{2025}).

\bibitem{Thampi2014}
\bibinfo{author}{Thampi, S.~P.}, \bibinfo{author}{Golestanian, R.} \&
  \bibinfo{author}{Yeomans, J.~M.}
\newblock \bibinfo{title}{Instabilities and topological defects in active
  nematics}.
\newblock \emph{\bibinfo{journal}{EPL (Europhysics Letters)}}
  \textbf{\bibinfo{volume}{105}} (\bibinfo{year}{2014}).

\bibitem{PhysRevLett.89.058101}
\bibinfo{author}{Aditi~Simha, R.} \& \bibinfo{author}{Ramaswamy, S.}
\newblock \bibinfo{title}{Hydrodynamic fluctuations and instabilities in
  ordered suspensions of self-propelled particles}.
\newblock \emph{\bibinfo{journal}{Phys. Rev. Lett.}}
  \textbf{\bibinfo{volume}{89}}, \bibinfo{pages}{058101}
  (\bibinfo{year}{2002}).

\bibitem{Calderer2025}
\bibinfo{author}{Calderer, M.~C.} \emph{et~al.}
\newblock \bibinfo{title}{Chevron patterns in an active nematic liquid crystal
  film in contact with smectic a}.
\newblock \emph{\bibinfo{journal}{Proceedings of the Royal Society A:
  Mathematical, Physical and Engineering Sciences}}
  \textbf{\bibinfo{volume}{481}} (\bibinfo{year}{2025}).

\bibitem{velez2024}
\bibinfo{author}{Vélez-Cerón, I.}, \bibinfo{author}{Guillamat, P.},
  \bibinfo{author}{Sagués, F.} \& \bibinfo{author}{Ignés-Mullol, J.}
\newblock \bibinfo{title}{Probing active nematics with in situ microfabricated
  elastic inclusions}.
\newblock \emph{\bibinfo{journal}{Proc. Natl. Acad. Sci. U S A}}
  \textbf{\bibinfo{volume}{121}}, \bibinfo{pages}{e2312494121}
  (\bibinfo{year}{2024}).

\bibitem{Blanc2023}
\bibinfo{author}{Blanc, C.} \emph{et~al.}
\newblock \bibinfo{title}{Helfrich-hurault elastic instabilities driven by
  geometrical frustration}.
\newblock \emph{\bibinfo{journal}{Reviews of Modern Physics}}
  \textbf{\bibinfo{volume}{95}} (\bibinfo{year}{2023}).

\bibitem{Lopez-Leon2012}
\bibinfo{author}{Lopez-Leon, T.}, \bibinfo{author}{Fernandez-Nieves, A.},
  \bibinfo{author}{Nobili, M.} \& \bibinfo{author}{Blanc, C.}
\newblock \bibinfo{title}{Smectic shells}.
\newblock \emph{\bibinfo{journal}{J Phys Condens Matter}}
  \textbf{\bibinfo{volume}{24}}, \bibinfo{pages}{284122}
  (\bibinfo{year}{2012}).

\bibitem{Lopez-Leon2011}
\bibinfo{author}{Lopez-Leon, T.}, \bibinfo{author}{Fernandez-Nieves, A.},
  \bibinfo{author}{Nobili, M.} \& \bibinfo{author}{Blanc, C.}
\newblock \bibinfo{title}{Nematic-smectic transition in spherical shells}.
\newblock \emph{\bibinfo{journal}{Physical Review Letters}}
  \textbf{\bibinfo{volume}{106}}, \bibinfo{pages}{247802}
  (\bibinfo{year}{2011}).

\bibitem{ramaswamy10}
\bibinfo{author}{Ramaswamy, S.}
\newblock \bibinfo{title}{The mechanics and statistics of active matter}.
\newblock \emph{\bibinfo{journal}{Annual Review of Condensed Matter Physics}}
  \textbf{\bibinfo{volume}{1}}, \bibinfo{pages}{323--345}
  (\bibinfo{year}{2010}).

\bibitem{marchetti2013}
\bibinfo{author}{Marchetti, M.~C.} \emph{et~al.}
\newblock \bibinfo{title}{Hydrodynamics of soft active matter}.
\newblock \emph{\bibinfo{journal}{Reviews of Modern Physics}}
  \textbf{\bibinfo{volume}{85}}, \bibinfo{pages}{1143--1189}
  (\bibinfo{year}{2013}).

\bibitem{Prost15}
\bibinfo{author}{Prost, J.}, \bibinfo{author}{Julicher, F.} \&
  \bibinfo{author}{Joanny, J.~F.}
\newblock \bibinfo{title}{Active gel physics}.
\newblock \emph{\bibinfo{journal}{Nature Physics}}
  \textbf{\bibinfo{volume}{11}}, \bibinfo{pages}{111--117}
  (\bibinfo{year}{2015}).

\bibitem{marenduzzo2007steady}
\bibinfo{author}{Marenduzzo, D.}, \bibinfo{author}{Orlandini, E.},
  \bibinfo{author}{Cates, M.} \& \bibinfo{author}{Yeomans, J.}
\newblock \bibinfo{title}{Steady-state hydrodynamic instabilities of active
  liquid crystals: Hybrid lattice boltzmann simulations}.
\newblock \emph{\bibinfo{journal}{Physical Review E}}
  \textbf{\bibinfo{volume}{76}}, \bibinfo{pages}{031921}
  (\bibinfo{year}{2007}).

\bibitem{C9SM00859D}
\bibinfo{author}{Coelho, R. C.~V.}, \bibinfo{author}{Araújo, N. A.~M.} \&
  \bibinfo{author}{Telo~da Gama, M.~M.}
\newblock \bibinfo{title}{Active nematic–isotropic interfaces in channels}.
\newblock \emph{\bibinfo{journal}{Soft Matter}} \textbf{\bibinfo{volume}{15}},
  \bibinfo{pages}{6819--6829} (\bibinfo{year}{2019}).

\bibitem{Sacha2018}
\bibinfo{author}{Sacha, K.} \& \bibinfo{author}{Zakrzewski, J.}
\newblock \bibinfo{title}{Time crystals: a review}.
\newblock \emph{\bibinfo{journal}{Rep Prog Phys}}
  \textbf{\bibinfo{volume}{81}}, \bibinfo{pages}{016401}
  (\bibinfo{year}{2018}).

\bibitem{Zalatel2023}
\bibinfo{author}{Zaletel, M.~P.} \emph{et~al.}
\newblock \bibinfo{title}{Colloquium: Quantum and classical discrete time
  crystals}.
\newblock \emph{\bibinfo{journal}{Reviews of Modern Physics}}
  \textbf{\bibinfo{volume}{95}}, \bibinfo{pages}{031001}
  (\bibinfo{year}{2023}).

\bibitem{Shapere2012}
\bibinfo{author}{Shapere, A.} \& \bibinfo{author}{Wilczek, F.}
\newblock \bibinfo{title}{Classical time crystals}.
\newblock \emph{\bibinfo{journal}{Phys Rev Lett}}
  \textbf{\bibinfo{volume}{109}}, \bibinfo{pages}{160402}
  (\bibinfo{year}{2012}).

\bibitem{Hyman1992}
\bibinfo{author}{Hyman, A.~A.}, \bibinfo{author}{Salser, S.},
  \bibinfo{author}{Drechsel, D.~N.}, \bibinfo{author}{Unwin, N.} \&
  \bibinfo{author}{Mitchison, T.~J.}
\newblock \bibinfo{title}{Role of gtp hydrolysis in microtubule dynamics:
  information from a slowly hydrolyzable analogue, gmpcpp.}
\newblock \emph{\bibinfo{journal}{Molecular Biology of the Cell}}
  \textbf{\bibinfo{volume}{3}}, \bibinfo{pages}{1155--1167}
  (\bibinfo{year}{1992}).

\bibitem{Ellis2020}
\bibinfo{author}{Ellis, P.~W.}, \bibinfo{author}{Nambisan, J.} \&
  \bibinfo{author}{Fernandez-Nieves, A.}
\newblock \bibinfo{title}{Coherence-enhanced diffusion filtering applied to
  partially-ordered fluids}.
\newblock \emph{\bibinfo{journal}{Molecular Physics}}
  \textbf{\bibinfo{volume}{118}}, \bibinfo{pages}{1--9} (\bibinfo{year}{2020}).

\bibitem{Thielicke2021}
\bibinfo{author}{Thielicke, W.} \& \bibinfo{author}{Sonntag, R.}
\newblock \bibinfo{title}{Particle image velocimetry for matlab: Accuracy and
  enhanced algorithms in pivlab}.
\newblock \emph{\bibinfo{journal}{Journal of Open Research Software}}
  \textbf{\bibinfo{volume}{9}} (\bibinfo{year}{2021}).

\bibitem{Thampi_2015}
\bibinfo{author}{Thampi, S.~P.}, \bibinfo{author}{Doostmohammadi, A.},
  \bibinfo{author}{Golestanian, R.} \& \bibinfo{author}{Yeomans, J.~M.}
\newblock \bibinfo{title}{Intrinsic free energy in active nematics}.
\newblock \emph{\bibinfo{journal}{Europhysics Letters}}
  \textbf{\bibinfo{volume}{112}}, \bibinfo{pages}{28004}
  (\bibinfo{year}{2015}).

\bibitem{velezceron2025activenematicpumps}
\bibinfo{author}{Vélez-Ceron, I.} \emph{et~al.}
\newblock \bibinfo{title}{Active nematic pumps} (\bibinfo{year}{2025}).
\newblock \urlprefix\url{https://arxiv.org/abs/2407.09960}.
\newblock
  \bibinfo{eprint}{{\href{https://arxiv.org/abs/2407.09960}{{arXiv:2407.09960}}}}.

\bibitem{PhysRevE.106.014705}
\bibinfo{author}{Saghatchi, R.}, \bibinfo{author}{Yildiz, M.} \&
  \bibinfo{author}{Doostmohammadi, A.}
\newblock \bibinfo{title}{Nematic order condensation and topological defects in
  inertial active nematics}.
\newblock \emph{\bibinfo{journal}{Phys. Rev. E}}
  \textbf{\bibinfo{volume}{106}}, \bibinfo{pages}{014705}
  (\bibinfo{year}{2022}).
\newblock \urlprefix\url{https://link.aps.org/doi/10.1103/PhysRevE.106.014705}.

\end{thebibliography}

\end{document}